\def\gte{\lower 0.5ex\hbox{${}\buildrel>\over\sim{}$}}
\def\lte{\lower 0.5ex\hbox{${}\buildrel<\over\sim{}$}}
\def\rho{{\it p}}

\documentstyle[12pt,aasms4,psfig]{article}

\begin{document}

\title{Gamma-rays from Galactic Black Hole Candidates
with Stochastic Particle Acceleration}

\author{Hui Li\altaffilmark{1,3}, Masaaki Kusunose\altaffilmark{2},
and Edison P. Liang\altaffilmark{1}}

\altaffiltext{1}{Department of Space Physics and Astronomy, Rice University,
Houston, TX 77251}
\altaffiltext{2}{Department of Astronomy, University of Texas,
Austin, TX 78712}
\altaffiltext{3}{NIS-2, MD436, Los Alamos National Lab., Los Alamos,
NM 87545, hli@lanl.gov}

\begin{abstract}

We consider stochastic particle acceleration in 
plasmas around stellar mass black holes to explain 
the emissions above 1 MeV from Galactic
black hole candidates.
We show that for certain parameter regimes, 
electrons can overcome Coulomb losses and be 
accelerated beyond the thermal distribution 
to form a new population, whose distribution
is broad and usually not a power law;
the peak energy of the distribution is determined by the 
balance between acceleration and cooling, 
with particles piling up around it.
Radiation by inverse Compton scattering
off the thermal (from background) and 
non-thermal (produced by acceleration) particles
can in principle explain the hard X-ray to gamma-ray 
emissions from black hole candidates.  
We present model fits of Cyg X-1 and GRO J0422
in 50 keV -- 5 MeV region observed with OSSE and COMPTEL.
\end{abstract}


\section{INTRODUCTION}

Most of Galactic black hole candidates (GBHCs)
show X-ray spectra well fit by a
thermal Comptonization model with temperatures $\sim$ 50 -- 100 keV
and Thomson depths of a few (e.g., \cite{har94}; \cite{lia93}).
OSSE and COMPTEL experiments on {\em Compton Gamma-Ray
Observatory} (CGRO), however, have recently revealed that persistent 
gamma-rays $>$ MeVs are being produced 
in some GBHCs, in particular Cyg X-1 
(\cite{joh93}; \cite{mcc94}) 
and GRO J0422+32 (\cite{dij95}).
These gamma-ray tails are hard to fit with a single component  
Sunyaev \& Titarchuk (1980) thermal model 
(or a more recent model by Titarchuk 1994).
Hence non-thermal processes are strongly hinted 
by these observations.

Models of non-thermal $e^\pm$ pairs have been studied 
with the emphasis on the power-law X-ray emission from 
active galactic nuclei (AGNs)
(e.g., \cite{fab86}; \cite{lz87}; \cite{sv87}; \cite{cop92}),
and they can be also applied to plasmas around 
stellar mass black holes.
Those models assume that either mono-energetic or power-law 
leptons with a large Lorentz factor ($\gamma \gg 1$) are injected
and they initiate cascade processes such as 
$e^{\pm}$ production and Compton scattering 
(see \cite{sv94} for a recent review). 
However, those models did not specify any acceleration mechanism,
thereby lacking the self-consistency in determining 
the particle distributions and photon spectra.

Motivated by the MeV photons from GBHCs, we
study the role of stochastic particle 
acceleration in accreting plasmas near GBHCs.
The mechanisms of wave-particle resonant interactions that lead to 
particle acceleration have been directly observed in solar-wind
(e.g., \cite{mar91}) and extensively
studied in the context of solar flares (e.g., \cite{mel74};
\cite{ram79}; \cite{hp92}; \cite{mr95}). 
Stochastic acceleration has been applied to 
diffusive shock acceleration and 
cosmic rays (\cite{sch94}), the lobes of radio galaxies 
(e.g., \cite{lac77}; \cite{ach79}; \cite{eil84}), 
and recently, the central regions of AGNs (\cite{dml95}).
Here in this {\em Letter},
we couple the particle energization and radiative
processes to determine self-consistently the steady state particle 
and photon distributions in GBHC environments.
We then present fits to gamma-ray
data of GBHCs observed with OSSE and COMPTEL.

\section{THE MODEL}
\label{model-sec}

As a first approximation, we average all plasma and magnetic 
field properties over an ad hoc spherical volume around the 
black hole of radius $R$. In this homogeneous spherical 
emission region, we treat the relevant radiation and particle 
acceleration/heating processes comprehensively, but ignore
details of the accretion flow, primary energy 
generation and viscosity, etc., by assuming naively that 
the total GBHC luminosity is released 
uniformly throughout the spherical volume.  
This approach allows us to first concentrate on 
the microphysics. 
We assume that the plasma cloud is in a steady state,
and solve the coupled steady-state kinetic equations of 
particles and photons (e.g., \cite{lz87}).
Our approach differs from previous studies in that
we include the {\em diffusion and systematic acceleration terms} 
in the particle kinetic equations, both of which are 
derived from the stochastic acceleration. We also assume 
that the particles consist of a thermal and non-thermal 
component, coupled by Coulomb interactions.


The evolution of the non-thermal particle distribution
in momentum space, $n_{\rm nt}(\rho)$, can be described by the
Fokker-Planck equation as
\begin{equation}
\label{fp-par}
{\partial n_{\rm nt} \over \partial t}  = 
\frac{\partial}{\partial \rho} \left[ D(\rho) \, 
\frac{\partial n_{\rm nt}}
{\partial \rho} \right] - \frac{\partial}{\partial \rho} \left[
\frac{2}{\rho} \, D(\rho) \, n_{\rm nt}
+ \dot{\rho} \, n_{\rm nt} \right] 
- \frac{n_{\rm nt}}{t_{\rm esc}(\rho)}
+ \dot{Q}_{\rm inj}(\rho) 
- \dot{A}(\rho)  \, ,
\end{equation}
where $\rho \equiv {|\vec {\rm p}|}/m_e c = \beta\gamma$,
$\gamma$ is particle's Lorentz factor,
and $\beta = (1-1/\gamma^2)^{1/2}$. 
Here, $D(\rho)$ is the particle diffusion coefficient 
in momentum space, $t_{\rm esc}$ is the energy-dependent escape timescale,
$\dot \rho$ is the cooling losses (both inverse Compton scattering 
and Coulomb losses are included), 
$\dot{Q}_{\rm inj}(\rho)$ is the injection
including sources from both the thermal background plasma 
and $e^{\pm}$ pairs production,
and $\dot{A}(\rho)$ represents the removal of leptons due to
pair annihilation ($e^+ + e^- \rightarrow \gamma + \gamma$).
The non-thermal particle density is given by
$\int n_{\rm nt}(\rho) d \rho$.
A direct consequence of including $D(\rho)$ 
(also the difference from previous study) is the
development of a population of non-thermal leptons 
which is accelerated out of the background thermal plasma.
However, below a certain momentum $\rho_{\rm thr}$, 
which we assume to correspond to the plasma temperature $T_e$
and $\gamma_{\rm thr} = (\rho_{\rm thr}^2 + 1)^{1/2}
=  1 + 4 \Theta$ (see e.g., \cite{zcl90}),
the particle distribution should be Maxwellian,
possibly due to either waves are strongly damped at those energies
or ``thermalization'' timescale (such as Coulomb collisions)
becomes the shortest.  Thus all the particles with
$\rho < \rho_{\rm thr}$ are grouped into the thermal population.
We concentrate on the steady state solutions 
($\partial / \partial t = 0$).

We admit that the actual distribution of the turbulence are 
subject to great uncertainty. We assume that turbulent
plasma waves are generated uniformly in the cloud and that
there is a power-law distribution of turbulence energy density 
in wave number $ W(k) \propto k^{-q}$, 
where $k = |{\vec k}|$, with $k$ extending from 
$k_{\rm min}$ (which is taken to be $\sim 1/R$) 
to the value near electron cyclotron resonance, 
so that it covers both the Alfv\'en and whistler branches.
Both $D(\rho)$ and $t_{\rm esc}$ can be evaluated from the
conditions of accreting plasma (see recent treatment in 
Dermer, Miller, \& Li, 1995). Here we combine the results from
both the electron/whistler and electron/Alfv\'en interactions,
and obtain the pitch-angle averaged momentum diffusion coefficient as 
\begin{equation}
\label{dpp-eq}
D(\rho)={\pi\over 4} \, (q-1) \, \frac{1}{t_{\rm dyn}} \,
{\beta_{\rm A}^2\over\beta} \, \zeta_{\rm wave} \,
\left( \frac{r_L}{R} \right)^{q-2}~I_a(k_0, k_1) \, \rho^q \, ,
\end{equation}
where
\[
I_a(k_0,k_1) = {\beta_{\rm A}^2 k_0^2 -1 \over q}
\left[ \left( {k_1\over k_0} \right)^{-q} -1 \right] 
+ {1\over q+2} \left[ \left( {k_1\over k_0} \right)^{-q-2} - 1 \right] 
+ {\beta_{\rm A}^2 \, k_0^2 \over 2-q}
\left[ \left( {k_1\over k_0} \right)^{2-q} -1 \right] 
\]
with $t_{\rm dyn} = R/c$, $\zeta_{\rm wave} = U_w / U_B$ 
(the ratio of wave to total magnetic energy density), 
$r_L = m_e c^2 / eB$ (the electron gyroradius),
$k_1 = (m_p / m_e)^{1/2} / \beta_A$, 
$k_0 = (m_p / m_e) / \rho$,
and $c \beta_{\rm A} = B/(4\pi n_p m_p)^{1/2}$
being the Alfv\'en velocity, where $n_p$ is the proton 
number density.  Here $B$ is obtained by assuming 
equipartition between thermal gas pressure and magnetic pressure,
and in most cases we study here, $\beta_{\rm A} \sim 0.01$.

The importance of stochastic acceleration can be illustrated
in Figure 1, where we plot the various timescales
(in units of $t_{\rm dyn}=R/c$) as functions of the dimensionless
particle momentum $\rho$ for a test electron passing through a
plasma cloud with size $R \approx 1.5\times 10^8$ cm,
temperature $\Theta \equiv kT_e/(m_ec^2) \approx 0.2$,
and varying density $n_p$.  Here we assume that
magnetic energy density $U_B$ is in equipartition with 
thermal gas pressure ($n_p\Theta$)
and $\zeta_{\rm wave} = 0.1$. 
Also, $u_{\rm ph} \equiv U_{\rm ph} /U_B$, where $U_{\rm ph}$ 
is the soft photon energy density. 
All timescales are evaluated from energy change rates. 
The acceleration rate is calculated from
$\langle d\gamma/dt \rangle_{\rm acc} = 
{1\over \rho^2} {\partial \over \partial \rho}\left[\rho^2 \beta
D(\rho)\right]$ and equation (\ref{dpp-eq}). 
The Compton+synchrotron cooling rate is
$|\langle d\gamma/dt \rangle_{\rm c+s}| = {4\over 3}
{1\over t_{\rm dyn}} (1+u_{\rm ph}) \tau_p \Theta \rho^2$,
where $\tau_p = \sigma_{\rm T} n_p R$
and $\sigma_{\rm T}$ is the Thomson cross section.
The cooling rate of lepton-lepton Coulomb interactions
is taken from Dermer \& Liang (1989).
The diffusive escape timescale is typically the longest among
all timescales and is not shown here.
It is evident that, for certain parameter regimes,
acceleration is more efficient than the cooling processes
so that it must be taken into account in determining the 
particle distributions of emitting plasma.


The energy flow rates for various components
in the system are outlined as follows:
(A) The total gravitational energy released by accretion is
characterized by the compactness $\ell$, which is
given by $\ell \equiv 4 \pi \sigma_{\rm T} R^2 \dot{u} / (3 m_e c^3)$,
where $\dot{u}$ is the energy input rate per unit volume;
(B) A fraction of gravitational energy,
$\ell_{\rm nt} = \epsilon_{\rm nt} \ell$,
is used to stochastically accelerate some leptons through
wave-particle resonant interactions; 
(C) The rest of the input energy, 
$\ell_{\rm th} = (1 - \epsilon_{\rm nt}) \ell$, 
is used to heat up thermal leptons; 
(D) Soft photons, with a blackbody spectrum of temperature $T_s$, 
are injected in the cloud uniformly with the 
compactness $\ell_{\rm s}$
(the source of soft photons can be 
optically thick accretion disks).

In determining the particle distribution and emission spectra,
we have included most of the relevant microphysical processes,
all of which are put into a computer code
(\cite{l95}; see also \cite{km95} and \cite{lz87}). 
These include
(A) {\em Non-thermal processes:} Compton cooling,
Coulomb collisions and annihilation between 
non-thermal and thermal leptons,
the flow of non-thermal pairs becoming 
thermal pairs due to cooling, and the escape;
(B) {\em Thermal Processes:}
Coulomb collisions between protons and thermal $e^{\pm}$, 
thermal bremsstrahlung and Compton scattering by thermal $e^{\pm}$, 
and annihilation emission by thermal $e^{\pm}$; 
Thermal Compton scattering is calculated using the Fokker-Planck equation
(Kompaneets equation) with the diffusion coefficient given
by Prasad et al. (1988) for isotropic radiation field.
When $\tau_{\rm e} = \sigma_{\rm T} n_e R < 2$ and $T_e > 100$ keV, 
however, we use a diffusion coefficient for the collimated 
radiation field, based on the argument by Titarchuk (1994).
(C) {\em Radiation Processes:} Additional processes include 
the injection of soft photons $\ell_s$, 
absorption due to $\gamma+\gamma$ interactions,
and the escape from the system;
(D) {\em Pair Balance:} Thermal pairs are assumed in pair balance, 
i.e., the pair production rate is equal to the annihilation rate 
for $\rho < \rho_{\rm thr}$, and escape is neglected.
 
To summarize our model, for given $\ell$, proton density
$n_p$, and $R$, distributions of steady state
particles and photons for a plasma cloud with the processes 
described above are obtained for given parameters such as 
$\epsilon_{\rm nt}$, $\ell_s$, $T_s$, and $q$.
The temperature of the plasma $T_e$ is mainly
determined by the balance among external heating (by protons)
$\ell_{\rm th}$, Coulomb heating by non-thermal leptons,
and cooling by thermal Compton scattering.
The Thomson scattering depth for thermal leptons 
$\tau_{\rm e} = \sigma_{\rm T} n_e R$
(usually different from $\tau_p = \sigma_{\rm T} n_p R$ due to pairs) 
is also obtained.  The level of
the turbulent plasma waves $\zeta_{\rm wave}$ is unambiguously
determined from the relation
$\ell_{\rm nt} = (4\pi\sigma_{\rm T}R^2/3c)
\int d\rho n_{\rm nt}(\rho) \langle{d\gamma/dt} \rangle_{\rm acc}$,
where $\langle{d\gamma/dt} \rangle_{\rm acc}$ is proportional
to $\zeta_{\rm wave}$ and $n_{\rm nt}(\rho)$ is obtained from equation
(\ref{fp-par}).
The advantage of this approach is that we do not need 
to artificially assume the injection of relativistic leptons as was 
done previously, because the content of non-thermal particles is now
self-consistently determined from the stochastic particle acceleration. 

\section{RESULTS} 
\label{resul-sec}

Figure \ref{fig2-rev.eps} summarizes our main results. 
Photon spectra of Cyg X-1 and GRO J0422
from OSSE (cross, \cite{phl95,kro95}) and COMPTEL
(filled square, \cite{mcc94b,dij95}) are shown in the left panels.

\noindent {\bf (1) Cyg X-1}.
Our best-fit model spectrum ({\em left}) to Cyg X-1 and the corresponding 
particle distribution ({\em right}) are shown as the solid curves
in the upper panel of Figure \ref{fig2-rev.eps}.
Other parameters are given in Table \ref{para-table}.
Clearly a deviation from Maxwellian 
of the particle distribution occurs at high energy tail.
This occurs because the ``upward'' motion (in momentum space)
by acceleration is balanced with the ``downward'' 
motion by cooling at a specific value of momentum, at which
particles tend to ``pile up'', forming a ``bump'' 
in the particle distribution (see also Schlickeiser 1984).
It is from this ``bump'' component that most of
the emissions above 1 MeV are produced via inverse 
Compton scattering.

Compared with the thermal models, only
$\ell_{\rm nt}$ and $q$ are the new 
additional parameters.  While the dependence of the results on 
$q$ is rather weak, we demonstrate the effects of varying 
$\ell_{\rm nt}$ in the upper panels of Figure \ref{fig2-rev.eps} 
(dotted and dashed curves).
Dotted curve ($\ell_{\rm nt}/\ell = 10^{-8}$)
indicates that photon spectrum from Cyg X-1 can not
be fitted by pure thermal emissions.
As $\ell_{\rm nt}$ increases, more pairs are produced (dashed curve
in particle distributions), so is $> MeV$ emission in this case (but
see below). Note that the pair fraction ($1-\tau_p/\tau_e$)
in this case is small ($\sim 3.4\%$).

\noindent {\bf (2) GRO J0422}.
A good fitting to GRO J0422 is shown by the solid curves
in lower panel of Figure \ref{fig2-rev.eps}.
For $M = 10 M_{\sun}$, we find that
it only needs $\ell_{\rm nt} / \ell = 0.06$,
and the particle distribution is dominated by the thermal 
component with only a slight deviation from the Maxwellian distribution
at higher energy due to acceleration.
This is also evident from the fact that
MeV emission from J0422 is much weaker, compared to Cyg X-1. 
We have also tried to fit J0422 by choosing a different mass
but keeping other parameters to be the same as for Cyg X-1
(as suggested by the referee). Dashed and dotted curves 
in the lower panel show other fittings 
(though not as good as the $M = 10 M_{\sun}$ case)
to J0422 with $M  = 6 M_{\sun}, \ell = 15$ 
and $M  = 4 M_{\sun}, \ell = 22.5$ (see Table \ref{para-table}). 
Note that the source luminosity is $\propto \ell~M$ for fixed $R$
so it is the same for all trials.
The particle distributions are now essentially thermal.
Again, other parameters are given in Table \ref{para-table}.

Figure \ref{fig2-rev.eps} also indicates an important 
feature of our model, i.e.  higher $\ell_{\rm nt} / \ell$ 
does not necessarily result in higher flux of gamma-ray emissions.
In the case of J0422, the high compactness 
(smaller mass) causes the cooling 
becoming increasingly important 
at high energies, destroying the otherwise suprathermal
population and merging them into thermal bath.  Notice that
the cutoff energy in photon spectra decreases and the number of pairs 
increases dramatically ($\sim 26\%$ for $4 M_{\sun}$) as mass decreases.

\section{CONCLUSIONS AND DISCUSSIONS}
\label{dis-sec}

We have examined stochastic particle acceleration
via wave-particle resonant interactions near Galactic black holes.
We find that under certain conditions, stochastic electron
acceleration can overcome both Coulomb and Compton losses,
resulting in a suprathermal population.
Preliminary model spectra show good fits to the recent OSSE 
and COMPTEL observations of Cyg X-1 and GRO J0422.
We find that Cyg X-1 has a much higher component in gamma-rays
than J0422, as it is also evident in the data.
This can be understood in terms of the higher
$\ell_{\rm nt} / \ell$ and lower compactness of Cyg X-1, 
compared to J0422, if they have the same black hole mass.
On the other hand, the high energy emissions
from J0422 can be fitted by different masses of the 
putative black holes.
But these fittings are different from the standard thermal 
(i.e. e-p plasma only) models because a large amount of
pairs are produced and thermalized due to the high
compactness. Since both sources are highly variable,
especially in gamma-ray emissions, we expect the
non-thermal energy content $\ell_{\rm nt} / \ell$ might
vary but the detailed microphysics of how the
system partitions the energy flows is presently unclear. 

The non-thermal population we obtain here is relatively soft with 
few pairs, not capable of explaining either the 
transient MeV bump discovered by Ling et al. (1987) or the 
COMPTEL power-law tail if it extends to much higher energies.  
This softness is due to our assumptions of a copious 
soft photon source permeating the entire emission region and 
{\em the coexistence of thermal and non-thermal 
particles in a single homogeneous volume}.  
Coulomb interaction {\em efficiently} transfers the 
non-thermal energy to the thermal reservoir which strongly
limits the existence and energy content of non-thermal particle
population.
To achieve much harder non-thermal populations and emissions 
we need to postulate the existence of physically distinct regions 
for the thermal and non-thermal components, either 
in the form of a thermal disk plus non-thermal corona or jets, 
or thermal outer disk and non-thermal inner torus.  
We plan to undertake such ventures in future work. But it is 
highly encouraging that a simple-minded single-component spherical 
model with few parameters is already capable of explaining 
from first principles the quiescent (low-hard state) 
emissions of Cyg X-1 and GRO J0422 detected by CGRO.  

We note, of course, that there are many 
other alternative interpretations of the hard tail, 
including the pion decay model (\cite{jr94}) and 
pair-dominated hot cloud model invoking quenching of the 
soft photon source (\cite{ld88}). It is possible that
particles accelerated by shocks may also account for the
$>$ MeV emissions.

\acknowledgments

We gratefully acknowledge the comments from an anonymous referee,
which helped to improve the manuscript.
We thank M. McConnell for providing the COMPTEL data for Cyg X-1 and
the OSSE team for both Cyg X-1 and GRO J0422 data. 
H.L. acknowledges the support of the Director's Postdoc Fellowship
at LANL and a Grant-in-Aid of Research from the 
National Academy of Sciences, through Sigma Xi, 
the Scientific Research Society. 
M.K. is grateful to J.C. Wheeler for his encouragement and 
acknowledges support by a grant from Texas Advanced Research Program. 
This work was also supported by NASA 
grants NAGW-2957, NAG 5-1547 and NRL contract No. N00014-94-P-2020.

\clearpage


\figcaption{
Timescales of acceleration ({\it solid}), 
Coulomb loss ({\it dotted}), and Compton loss ({\it dashed}) 
as functions of particle's dimensionless
momentum $\rho \equiv \gamma \beta$. 
The scattering depth and soft photon energy density, indicated respectively
by $\tau_p$ and $u_{\rm ph}$ (see text for definitions)
for a given size $R$, are varied as 
$(\tau_p, u_{\rm ph})~=~(0.05,1), ~(0.5,1),~
(0.5,10),$ and $(2,0.1)$, for plots 
(a), (b), (c) and (d), respectively.
Sufficient acceleration occurs in plots (a) and (b),
but Compton/syn. cooling and Coulomb cooling prevent acceleration
in (c) and (d), respectively.
\label{time.ps}
}

\figcaption{
{\it Upper panels}: Model fit of the emission spectrum of 
Cyg X-1 ({\em left}) and the corresponding particle distribution
as function of dimensionless momentum $p$ ({\em right}). 
Data are from OSSE and COMPTEL ({\it squares}) 
experiments. Solid curve ($\ell_{\rm nt}/\ell = 0.15$) fits the data.
Dotted and dashed curves are for $\ell_{\rm nt}/\ell = 10^{-8}$ 
and $0.5$, respectively.  They depict the effects of 
changing $\epsilon_{\rm nt}$ {\em only}, with other parameters fixed.
Particle distributions for higher $\ell_{\rm nt}/\ell$ (solid and dashed 
curves) show deviations from Maxwellian distributions
in the high energy tail due to acceleration. 
{\it Lower panels}: Same as the upper panels but for GRO J0422.
Again, data are from OSSE and COMPTEL.
The solid, dashed and dotted curves are for $M/M_{\sun} = 
10,~6$, and 4, respectively. Solid curve fits the data best
with $\ell_{\rm nt} / \ell =  0.06$, and the corresponding 
particle distribution is very close to Maxwellian. 
Particle distributions for $M/M_{\sun} = 6,~4$ are
essentially thermal and a large amount of pairs are produced
(dashed and dotted curves).
\label{fig2-rev.eps}
}

\clearpage

\begin{deluxetable}{ccccccccccc|ccc}
\footnotesize
\tablecaption{Model Parameters and Results \tablenotemark{a}
\label{para-table}}
\tablewidth{0pt}
\tablehead{
\colhead{Object} & \colhead{dist.} 
& \colhead{mass} & \colhead{$\ell$} 
& \colhead{$\ell/\ell_{\rm edd}$} & \colhead{$\ell_{\rm nt}/\ell$} 
& \colhead{$\ell_{s}/\ell$} & \colhead{$T_s$}
& \colhead{$\tau_p$} & \colhead{$q$} & \colhead{$R$} 
& \colhead{$T_e$} & \colhead{$\tau_e$} 
& \colhead{$\zeta_{\rm wave}$} 
}

\startdata

Cyg X-1\tablenotemark{b} & 2.5 & 10 & 4.5 & 0.02 & 0.15 & 0.07 
& 0.1 & 0.7 & $5/3$ & 80 & 139 & 0.725 & 0.059 \nl
J0422\tablenotemark{c} & 2.5 & 10 & 9.0 & 0.03 & 0.06 & 0.08 & 
0.1 & 0.7 & $5/3$ & 80 & 134 & 0.741 & 0.065 \nl
J0422\tablenotemark{d} & 2.5 & 6 & 15 & 0.05 & 0.15 & 0.07 & 0.1 
& 0.7 & $5/3$ & 80 & 119 & 0.869 & 0.08 \nl
J0422\tablenotemark{e} & 2.5 & 4 & 22.5 & 0.08 & 0.15 & 0.07 & 0.1 
& 0.7 & $5/3$ & 80 & 111 & 0.944 & 0.07 \nl
\enddata

\tablenotetext{a}{distance in kpc, mass in $M_{\sun}$,
temperature in keV, $R$ in units of $GM/c^2$}
\tablenotetext{b}{parameters for the solid curve for Cyg X-1 in
Figure \ref{fig2-rev.eps}}
\tablenotetext{c,d,e}{parameters for the solid, dashed and dotted curves 
for J0422 in Figure \ref{fig2-rev.eps}, respectively}

\end{deluxetable}


\clearpage

\begin{thebibliography}{}
\bibitem[Achterberg 1979]{ach79} 
 Achterberg, A. 1979, A\&A, 76, 276

\bibitem[Coppi 1992]{cop92} 
 Coppi, P. S. 1992, \mnras, 258, 657

\bibitem[Dermer \& Liang 1988]{dl89} 
 Dermer, C. D., \& Liang, E. P. 1989, \apj, 339, 512 

\bibitem[Dermer, Miller, \& Li 1995]{dml95} 
 Dermer, C. D., Miller, J. A., \& Li, H. 1995, \apj, in press

\bibitem[Eilek \& Henrikson 1984]{eil84} 
 Eilek, J. A., \& Henrikson, R. N. 1984, \apj, 277, 820

\bibitem[Fabian et al. 1986]{fab86} 
 Fabian, A. C., Blandford, R. D., Guilbert, P. W., Phinney, E. S., \&
 Cuellar, L. 1986, \mnras, 218, 171

\bibitem[Hamilton \& Petrosian 1992]{hp92} 
 Hamilton, R. J., \& Petrosian, V. 1992, \apj, 398, 350

\bibitem[Harmon et al. 1994]{har94} 
 Harmon, B. A. et al. 1994, in The Second Compton Symposium, 
ed. C. Fichtel, N. Gehrels, \& J. Norris 
(New York: AIP Conf. Proc. No. 304), 210

\bibitem[Johnson et al. 1993]{joh93} 
 Johnson, W. N. et al. 1993, \aaps, 97, 21

\bibitem[Jourdain \& Roques 1994]{jr94} 
 Jourdain, E., \& Roques, J. P. 1994, in The Second Compton Symposium, 
 ed. C. Fichtel, N. Gehrels, \& J. Norris 
 (New York: AIP Conf. Proc. No. 304), 329


\bibitem[Kroeger et al. 1995]{kro95} 
 Kroeger, R. A. et al. 1995, to be submitted

\bibitem[Kusunose \& Mineshige 1995]{km95} 
 Kusunose, M., \& Mineshige, S. 1995, \apj, 440, 100

\bibitem[Lacombe 1977]{lac77} 
Lacombe, C. 1977, A\&A, 54, 1

\bibitem[Li 1995]{l95} 
Li, H. 1995, PhD dissertation, Rice University

\bibitem[Liang 1993]{lia93}  
 Liang, E. P. 1993, in The first Compton Symposium, ed. M. Friedlander, 
 N. Gehrels \& D. Macomb (New York: AIP Conf. Proc. No. 280), 396

\bibitem[Liang \& Dermer 1988]{ld88} 
 Liang, E.P., \& Dermer, C.D. 1988, \apj, 325, L39


\bibitem[Lightman \& Zdziarski 1987]{lz87} 
 Lightman, A. P., \& Zdziarski A. A. 1987, \apj, 319, 643

\bibitem[Ling et al. 1987]{ling87}  
 Ling, J. C. et al. 1987, \apj, 321, L117

\bibitem[Marsch 1991]{mar91} 
 Marsch, E. 1991, in Physics of the Inner Heliosphere II,
 ed. R. Schwenn \&  E. Marsch (Springer-Verlag: Berlin)

\bibitem[McConnell et al. 1994a]{mcc94} 
 McConnell, M. et al. 1994, \apj, 424, 933

\bibitem[McConnell et al. 1994b]{mcc94b} 
 McConnell, M. et al. 1994, COSPAR meeting

\bibitem[Melrose 1974]{mel74} 
 Melrose, D. B. 1974, Sol. Phys., 37, 353

\bibitem[Miller \& Roberts 1995]{mr95} 
 Miller, J. A., \& Roberts, D. A. 1995, \apj, 452, 912

\bibitem[Phlips et al. 1995]{phl95} 
 Phlips, B. F. et al. 1995, \apj, {\it to be submitted}

\bibitem[Prasad et al. 1988]{pra88}
 Prasad, M. K. et al. 1988, \qjras, 40, 29

\bibitem[Ramaty 1979]{ram79} 
 Ramaty, R. 1979, in Particle Acceleration Mechanisms in Astrophysics,
 ed. J. Arons, C. Max, \& C. McKee (New York, AIP), 135

\bibitem[Schlickeiser 1984]{sch84} 
 Schlickeiser, R. 1984, A\&A, 136, 227

\bibitem[Schlickeiser 1994]{sch94} 
 Schlickeiser, R. 1994, ApJS, 90, 929



\bibitem[Sunyaev \& Titarchuk 1980]{st80}  
 Sunyaev, R., \& Titarchuk, L. 1980, \aap, 86, 121


\bibitem[Svensson 1994]{sv94} 
 Svensson, R. 1994, ApJS, 92, 585

\bibitem[Svensson 1987]{sv87} 
 Svensson, R. 1987, \mnras, 227, 403

\bibitem[Titarchuk 1994]{tit94} 
 Titarchuk, L. 1994, \apj, 434, 570

\bibitem[van Dijk et al. 1995]{dij95} 
 van Dijk, R. et al. 1995, A\&A, 296, L33

\bibitem[Zdziarski et al. 1990]{zcl90} 
 Zdziarski, A. A., Coppi, P. S., \& Lamb D. Q. 1990, \apj, 357, 149

\end{thebibliography}
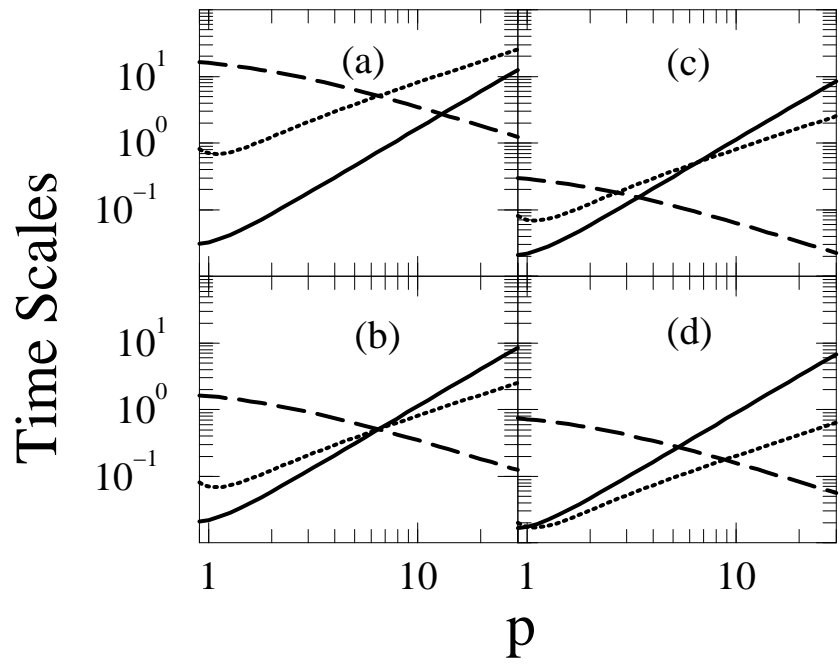
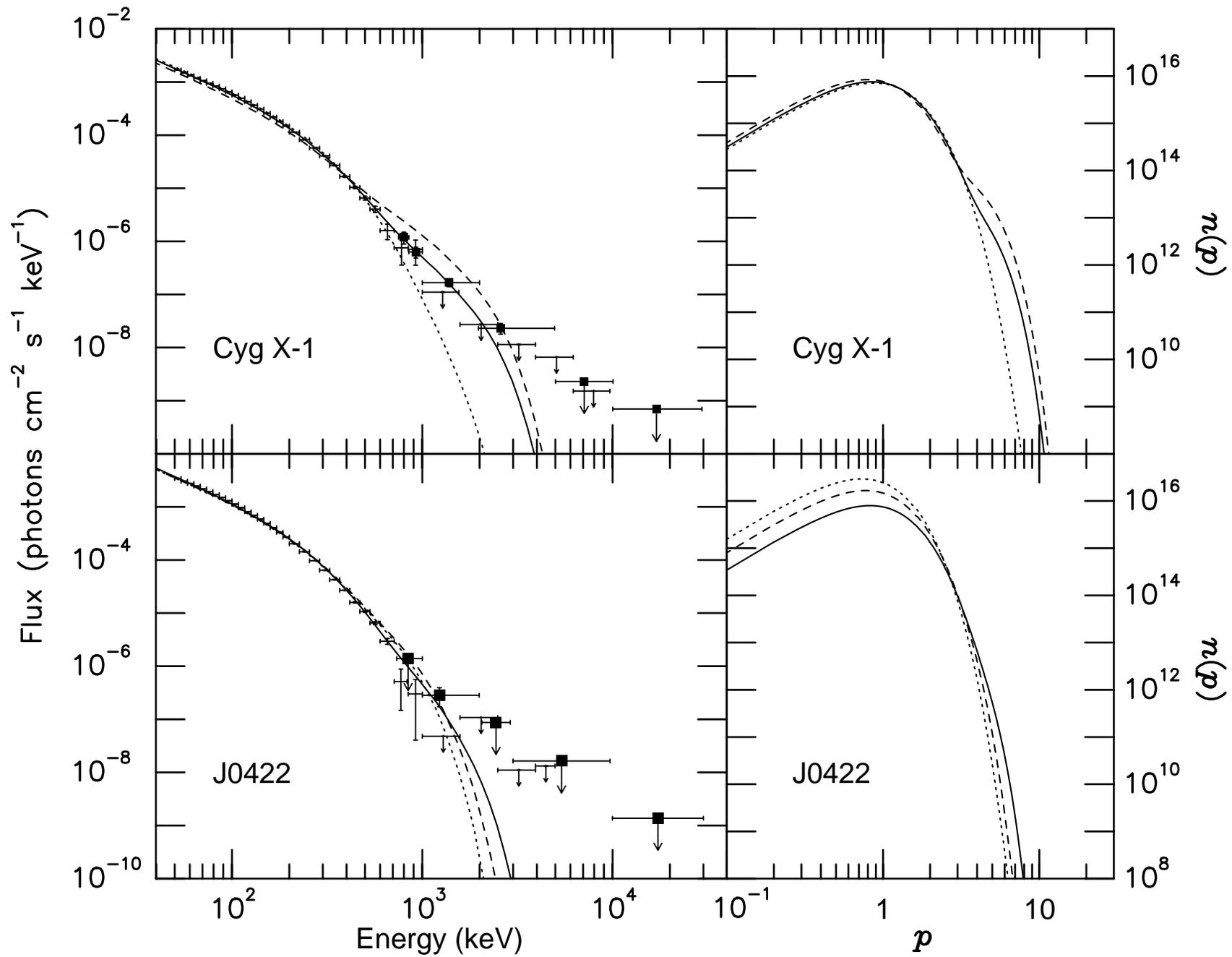
\end{document}